\documentclass[twocolumn]{aastex631}
\usepackage{tabularx,booktabs}
\usepackage{graphicx}

\usepackage{CJKutf8}
\newcommand{\chinesename}{{\begin{CJK}{UTF8}{gbsn}(王兰萱)\end{CJK}}}

\begin{document}

\title{Rescuing Kepler's False Positives: A Possible Habitable Zone Exoplanet in the Triple Star System KOI-1623}

\author[0000-0003-1953-3563]{Rita Lanxuan Wang \chinesename}
\affiliation{Institute for Astronomy, University of Hawai\textquoteleft i, 2680 Woodlawn Dr, Honolulu, HI 96822, USA}
\author[0000-0001-8832-4488]{Daniel Huber}
\affiliation{Institute for Astronomy, University of Hawai\textquoteleft i, 2680 Woodlawn Dr, Honolulu, HI 96822, USA}
\author[0000-0003-3244-5357]{Daniel Hey}
\affiliation{Institute for Astronomy, University of Hawai\textquoteleft i, 2680 Woodlawn Dr, Honolulu, HI 96822, USA}
\author[0009-0000-2051-8120]{Nathanael Burns-Watson}
\affiliation{Department of Astronomy, University of Texas at Austin, Austin, TX 78712, USA}
\author{Alexander Larsen}
\affiliation{Department of Astronomy, University of Texas at Austin, Austin, TX 78712, USA}
\author{Adam Kraus}
\affiliation{Department of Astronomy, University of Texas at Austin, Austin, TX 78712, USA}
\author{Trent J. Dupuy}
\affiliation{Institute for Astronomy, University of Edinburgh, Royal Observatory, Blackford Hill, Edinburgh, EH9 3HJ, UK}
\author[0000-0002-8965-3969]{Steven Giacalone}
\altaffiliation{NSF Astronomy and Astrophysics Postdoctoral Fellow}
\affiliation{Department of Astronomy, California Institute of Technology, Pasadena, CA 91125, USA}
\author{Justin R. Crepp}
\affiliation{University of Notre Dame, Department of Physics and Astronomy, Notre Dame, IN, USA}

\begin{abstract}
The \textit{Kepler} mission has significantly advanced our understanding of exoplanet demographics. However, \textit{Kepler} planet candidates showing centroid shifts are often classified as false-positives and thus excluded from demographic analyses. Some of these systems may be erroneous false positives, resulting from unresolved binary stars with planets orbiting one of the stars, thereby introducing biases into past exoplanet demographic studies. Here, we present updated \textit{Kepler} centroid offset values derived from \textit{Gaia} Data Release 3 (DR3) and analyze KOI-1623.01, a \textit{Kepler} centroid offset false positive (COFP) with a clear transit signal and an orbital period of 111 days. High-contrast imaging with Keck/NIRC2 reveals that KOI-1623A has a close companion with a separation of $0\farcs704 \pm 0\farcs002$, consisting of a slightly evolved G-type primary and a K-dwarf secondary, while \textit{Gaia} detects a tertiary companion with common proper motion at a separation of $\sim 8\farcs4$. By combining transit photometry, \textit{Kepler} pixel level data, direct imaging, and \textit{Gaia} DR3, we find that the transit signal is consistent with a $10.02^{+0.44}_{-0.31}\,R_{\earth}$ giant planet located in the habitable zone of the secondary. We furthermore use resolved companions from \textit{Gaia} to estimate that another $\approx$70 out of 515 of \textit{Kepler}’s centroid false positives were misclassified and instead are real planets that belong to binary systems.
\end{abstract}

\section{Introduction} \label{sec:context}

The \textit{Kepler} mission has transformed our understanding of exoplanet demographics \citep{Borucki2010Sci...327..977B, Koch2010ApJ...713L..79K}, confirming more than 2,000 exoplanets and revealing how stellar and planetary properties influence planet occurrence rates \citep[e.g.][]{Howard2012ApJS..201...15H, Mulders2015ApJ...814..130M, Fulton2017AJ....154..109F, chen2022AJ....163..249C}. However, most exoplanet demographic studies focus on single-star systems, while binary systems are common among solar-like star populations \citep{1991A&A...248..485D, 2008ApJ...679..762K,Raghavan2010ApJS..190....1R, 2011ApJ...731....8K}. Binary companions influence planet formation by truncating disks, exciting planetesimals' eccentricity and velocities, and enhancing accretion and photoevaporation \citep{Artymowicz1994ApJ...421..651A, Quintana2007ApJ...660..807Q, Alexander2012ApJ...757L..29A}. \citet{2016AJ....152....8K} showed that binarity suppresses planet occurrence rates among \textit{Kepler} planet candidates, highlighting a deficit of planet hosts in close binary systems. \citet{Ziegler2020AJ....159...19Z, Ziegler2021AJ....162..192Z} extended this analysis to TESS planet candidates and reached a similar conclusion. This broad result has also been demonstrated by \citet{Hirsch2021AJ....161..134H}, \citet{Fontanive2024ESS.....561403F}, and \citet{Thebault2025arXiv250618759T}.

Binarity also introduces observational biases that affect planet occurrence rates. For example, in the \textit{Kepler} mission, transit signals from binary systems can lead to false rejection of planetary candidates. The \textit{Kepler} catalog classifies transit signals as either planet candidates or false positives based on transit photometry. One such false positive flag is the centroid offset flag, which rejects targets with significant centroid offsets between the transit source and the target star, indicating that the transit source is occulting a different star \citep{Bryson2013PASP..125..889B, vetting}. Some of these are true false positives caused by background eclipsing binaries unrelated to the target star. However, real planets can also be rejected if a transiting planet occults the target’s binary companion. 

Many studies have identified planets in binary systems in S-type orbits among \textit{Kepler} targets. For example, \textit{Kepler}-296 is a binary system composed of two M-dwarfs separated by $0\farcs2$, and it hosts five planets that all orbit the primary star \citep{Barclay2015ApJ...809....7B}. \citet{Cartier2015ApJ...804...97C} used high-resolution imaging from the \textit{Hubble Space Telescope} to study \textit{Kepler} planet candidate hosts, focusing on three systems that were resolved as multi-star systems and providing updated stellar properties for each. \citet{2025arXiv250814176Z} reported two confirmed Earth-sized planets and one additional Earth-sized planet candidate in an M-dwarf binary system using TESS data. \citet{Zhang2025arXiv250925332Z} studied the orbital architecture of S-type planets in binary systems and found a bimodal distribution of planet–binary mutual inclinations, with one population being nearly aligned and the other exhibiting more scattered inclinations.

Approximately 450 KOIs have been flagged as false positives due to centroid offsets in Exoplanet Archive's DR25 KOI list \citep{koidr25}, and $\sim500$ in the Cumulative KOI list \citep{KOIcumulative}, which is a single catalog that compiles the KOI dispositions and transit information from multiple Data Release activity tables \citep{Akeson2013PASP..125..989A,Batalha2013ApJS..204...24B,Burke2014ApJS..210...19B,Rowe2015ApJS..217...16R,Mullally2015ApJS..217...31M,Thompson2016ksci.rept....9T,Coughlin2016ApJS..224...12C,Thompson2018ApJS..235...38T}. Although the absolute number of centroid offset false positives has increased across data releases, the percentage of KOIs flagged solely for centroid offsets has remained relatively stable ($\approx $5\%). No previous exoplanet demographic studies accounted for these cases.  

Re-examining these systems is important for two reasons. First, it refines our understanding of survey completeness. For instance, in the \textit{Kepler} sample, occurrence rates for long-period planets remain uncertain \citep{Foreman-Mackey2016AJ....152..206F, Herman2019AJ....157..248H, Bryson2021AJ....161...36B}. By investigating how many real planets may have been misclassified as false positives, we can better assess the impact of such misclassifications on completeness estimates. Second, identifying transiting planets in binary systems can improve our understanding of planetary properties and dynamics in these systems. For example, studies of planet-hosting binaries have shown that the stellar and planetary orbits generally have low mutual inclinations \citep{Dupuy2022MNRAS.512..648D, Christian2022AJ....163..207C, Behmard2022AJ....163..160B, Lester2023AJ....166..166L}, whereas \citet{Evans2024MNRAS.534..575E} reported that planet-hosting triple-star systems are typically not fully coplanar.

In this paper, we present a re-calculation of the \textit{Kepler} centroid offset values for all \textit{Kepler} targets using \textit{Gaia} astrometric data. Then, we focus our analysis on a particularly interesting target: KOI-1623.01, a \textit{Kepler} centroid-offset false positive with a clear transit signal in a triple-star system, featuring a close inner binary resolved by adaptive optics (AO) imaging. We perform a detailed analysis of the system and discuss broader implications for exoplanet demographics, including how many similar cases may have been overlooked.

\section{Updating \textit{Kepler} Centroid Offsets with \textit{Gaia} DR3}\label{sec:Gaia}
\subsection{\textit{Kepler} Centroid Offsets}
\textit{Kepler} false positives are classified into four main categories: Not Transit-Like (NT), Stellar Eclipse (SS), Centroid Offset (CO), and Ephemeris Match Indicates Contamination (EC). These classifications are assigned by an automated vetting algorithm called the Robovetter \citep{Thompson2018ApJS..235...38T}. The false-positive dispositions are also manually reviewed by the \textit{Kepler} False Positive Working Group (FPWG) and documented in the Certified False Positive (CFP) table \citep{Bryson2017ksci.rept...12B}. KOIs are classified as certified false positives based on the logic analysis described in \citet{Bryson2017ksci.rept...12B}. In this work, we focus on centroid offset false positives (COFP). 

\textit{Kepler}’s pixel scale is $3\farcs98$ \citep{Koch2010ApJ...713L..79K}, and typical photometric apertures span a radius of a few pixels \citep{Bryson2010SPIE.7740E..1DB}. Consequently, flux from nearby stars both within and outside the aperture can contaminate the target star’s signal, possibly mimicking a planetary transit signal on the target star. Such contamination is often flagged as a COFP by the Centroid Robovetter \citep{Mullally2017ksci.rept...23M}, indicating that the transit signal originates from a different source. Several astrophysical phenomena can produce false transit signals that result in centroid offsets. For instance, they can be caused by eclipsing binaries or large transiting planets located within or near the target star's aperture. Flux contamination can also arise from instrumental effects, such as optical ghosts or electronic crosstalk from distant transiting sources on the detector \citep{Caldwell2010ApJ...713L..92C}.

Multiple techniques are used to determine if a transit signal originates from a background source \citep{Bryson2013PASP..125..889B}. The difference imaging technique determines the transit source by subtracting out-of-transit from in-transit pixel values and fitting the resulting image to the Pixel Response Function (PRF) to derive centroid offsets \citep{Bryson2010ApJ...713L..97B, Bryson2013PASP..125..889B}. The PRF is fitted to both the difference image (the average out-of-transit image minus the in-transit image, which produces a star image at the location of the transit signal) and the out-of-transit (OOT) image (where the signal appears at the flux centroid of the target star) to determine the OOT centroid offset. The \textit{Kepler} Input Catalog (KIC; \citealp{KIC, kepler_KIC}) provides photometric data for stars in \textit{Kepler}’s field of view based on ground-based observations, with a median image full width at half maximum (FWHM) of $2\farcs5$. During the vetting process, the transit source position is also compared with the target's KIC position to calculate the KIC offset. In general, a system is marked as a COFP if the brightest pixel in the difference image is located more than 1.5 pixels ($\sim 6 \arcsec$) away from the target star's OOT centroid position. Contamination can also be unresolved from the target; in such cases, the system is flagged as a COFP if OOT centroid offset exceeds $2\arcsec$ at the $3\sigma$ significance level or exceeds $1\arcsec$ at the $4\sigma$ level \citep{Bryson2013PASP..125..889B,Thompson2018ApJS..235...38T}.

\subsection{Updated Centroids with \textit{Gaia} DR3}
\begin{figure*}
    \centering
    \includegraphics[width=\linewidth]{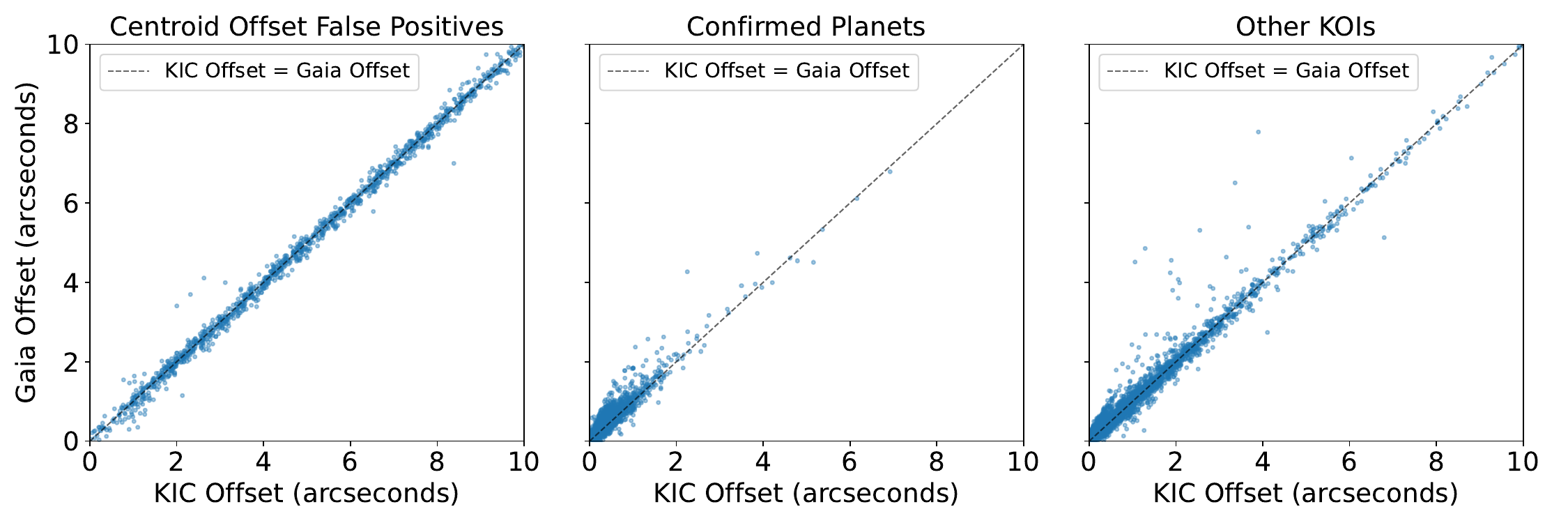}
    \caption{Comparison of centroid offsets for KIC targets derived using \textit{Gaia} and KIC positions. The three panels correspond to COFPs (left), confirmed planets (middle), and all other KOIs (right). The gray dashed line marks the one-to-one relation.}
    \label{fig:Gaia-KIC}
\end{figure*}

Because the original \textit{Kepler} centroid offsets were calculated before \textit{Gaia} data became available, we recompute the \textit{Kepler} Input Catalog (KIC) offsets for all KOIs using \textit{Gaia} Data Release 3 (DR3). \textit{Gaia} offers a significantly improved angular resolution of $\sim0\farcs7$ compared to the KIC’s $\sim2\farcs5$ \citep{Gaia2021A&A...649A...1G, KIC, GaiaDR3}. The out-of-transit (OOT) offsets, by contrast, are derived solely from \textit{Kepler} pixel-level data and are not recomputed here. The OOT offsets quoted throughout this work refer to the DR25 values. We note that the OOT offset measures the target's position from a PRF fit that assumes the target is a single star. For an unresolved binary, the fitted position is the flux-weighted photocenter of the blended components rather than either individual star, which biases the offset calculation \citep{Bryson2010ApJ...713L..97B, Bryson2013PASP..125..889B}.

We first cross-match all \textit{Gaia} targets within $4''$ of each KIC position. Due to \textit{Gaia}'s high angular resolution, a single KIC target may correspond to multiple \textit{Gaia} sources. We examined the distribution of $G - K$ colors for the cross-matched sources and found it to be bimodal, with peaks near $\sim\,1.5$ mag and $\sim \, 6.5$ mag. We applied a color constraint of $G - K < 2$ mag to exclude mismatched \textit{Gaia} sources, removing approximately half of the initially matched sample. This ensures that the \textit{Gaia} source is of comparable brightness and likely the intended target, rather than a background star.

The \textit{Gaia} centroid offset, defined as the angular separation between the \textit{Gaia} position of a source and the transit location, is calculated using the \textit{Gaia} coordinates, the KIC coordinates, and the reported KIC centroid offset. To derive the \textit{Gaia} offset, we first compute the difference between the target's \textit{Gaia} DR3 and KIC positions in projected right ascension and declination, add this correction to the reported KIC offset components, and then calculate the magnitude of the resulting two-dimensional vector.
Figure~\ref{fig:Gaia-KIC} compares the KIC centroid offsets with the \textit{Gaia}-based offsets computed in this work. 

The KIC and \textit{Gaia} centroid offsets show strong agreement, with residuals (KIC Offset - \textit{Gaia} Offset) having a mean of $-0\farcs0047\pm0\farcs0002$ and a standard deviation of $0\farcs084$. This agreement demonstrates that substituting \textit{Gaia} for KIC positions introduces no systematic bias.

Given the agreement, the outliers where the KIC and \textit{Gaia} offsets differ substantially are of particular interest. For example, a system with a small \textit{Gaia} offset but a large KIC offset may indicate that \textit{Gaia} has resolved the target into multiple stars, allowing the transit source to be localized on one of them. Conversely, systems with small offsets ($\lesssim2''$) in both catalogs may still contain unresolved companions that cause the offsets. Detecting such companions requires high-contrast imaging, motivating our detailed investigation of KOI-1623 below. 

We visually inspected the Data Validation (DV; \citealp{Twicken2018PASP..130f4502T,Li2019PASP..131b4506L}) Reports for systems with differing Gaia and KIC offsets and found that most involve stars in crowded fields, where KIC offsets are known to be less reliable \citep{vetting}. Figure~\ref{fig:Gaia-KIC} also shows that centroid offset values tend to be larger for COFPs, while confirmed planets exhibit much smaller offsets, typically $\lesssim2''$. 
For all COFP KOIs with close-in companions detected in the direct imaging surveys of \citet{2016AJ....152....8K} and \citet{Pbinpaper}, we visually inspected the DV reports of those with centroid offsets smaller than $2\arcsec$. We identified an interesting false positive target, KOI-1623.01, which exhibits a clear transit signal, along with OOT and KIC centroid offsets of $\sim 1\arcsec$. In the following section we present a detailed analysis of KOI-1623.01 as a case study of a possible genuine planet that has been misclassified as a COFP.

\section{Assessing the Nature of KOI-1623.01}

\subsection{Observations} \label{sec:data}

\subsubsection{\textit{Kepler} Photometry}
The KOI-1623 system was observed by \textit{Kepler} during Quarters 1-7, 9-11, 13-15, and 17, with a transit signal identified at a period of $P = 110.9$ days. The target is currently designated as a false positive (FP), with the only flag being the CENT\_UNRESOLVED\_OFFSET flag, indicating that it is classified as a COFP and that the transit signal does not originate from a resolved source. KOI-1623.01's false positive disposition was also manually reviewed by the \textit{Kepler} False Positive Working Group and certified as a false positive based on its centroid offsets, following the procedure described in \citet{Bryson2017ksci.rept...12B}. The transit signal shows a $1\farcs29$ offset relative to the out-of-transit centroid and a $1\farcs07$ offset relative to its KIC position. Our re-calculation using \textit{Gaia} astrometry yields a $0\farcs98$ offset relative to the \textit{Gaia} position. 

High-resolution imaging resolved KOI-1623 into a binary pair (Section~\ref{sec:nirc2}), which is unresolved in the \textit{Kepler} pixels. This has two consequences for the centroid diagnostics. First, the OOT centroid is derived from a single-star PRF fit and therefore reports the flux-weighted photocenter of the blended binary rather than either individual star, so the $1\farcs29$ OOT offset is biased and should be treated as approximate. Second, we visually inspected the per-quarter difference and OOT images in the DV report and find that the difference-image source is consistently offset from the OOT centroid in nearly all quarters with a detected transit. This systematic offset is what drives the COFP disposition. Because the disposition assumes a single star, it breaks down for an unresolved binary. The offset does not argue against a transiting planet but instead reflects the binarity of the system. Our pixel-level analysis in Section~\ref{sec:pixel}, which explicitly models both stellar components, confirms this interpretation.

We downloaded \textit{Kepler} light curves from the Mikulski Archive for Space Telescopes (MAST) through the \texttt{lightkurve} package \citep{lightkurve, kepler_MAST}. For KOI-1623, only long-cadence data (1800\,s) are available. We use the Presearch Data Conditioning Simple Aperture Photometry (PDC-SAP) product, which provides flux measurements within the optimal aperture after correcting for systematic effects from the telescope and spacecraft \citep{Thompson2016ksci.rept....9T}. All available light curves are stitched and normalized using \texttt{lightkurve}. We then remove \texttt{NaN} values and apply sigma clipping to filter out outliers, using an upper threshold of $5\sigma$ and a lower threshold of $10\sigma$.

\subsubsection{High Resolution Imaging with Keck/NIRC2} \label{sec:nirc2}
\begin{figure}
  \includegraphics[width=0.47\textwidth]{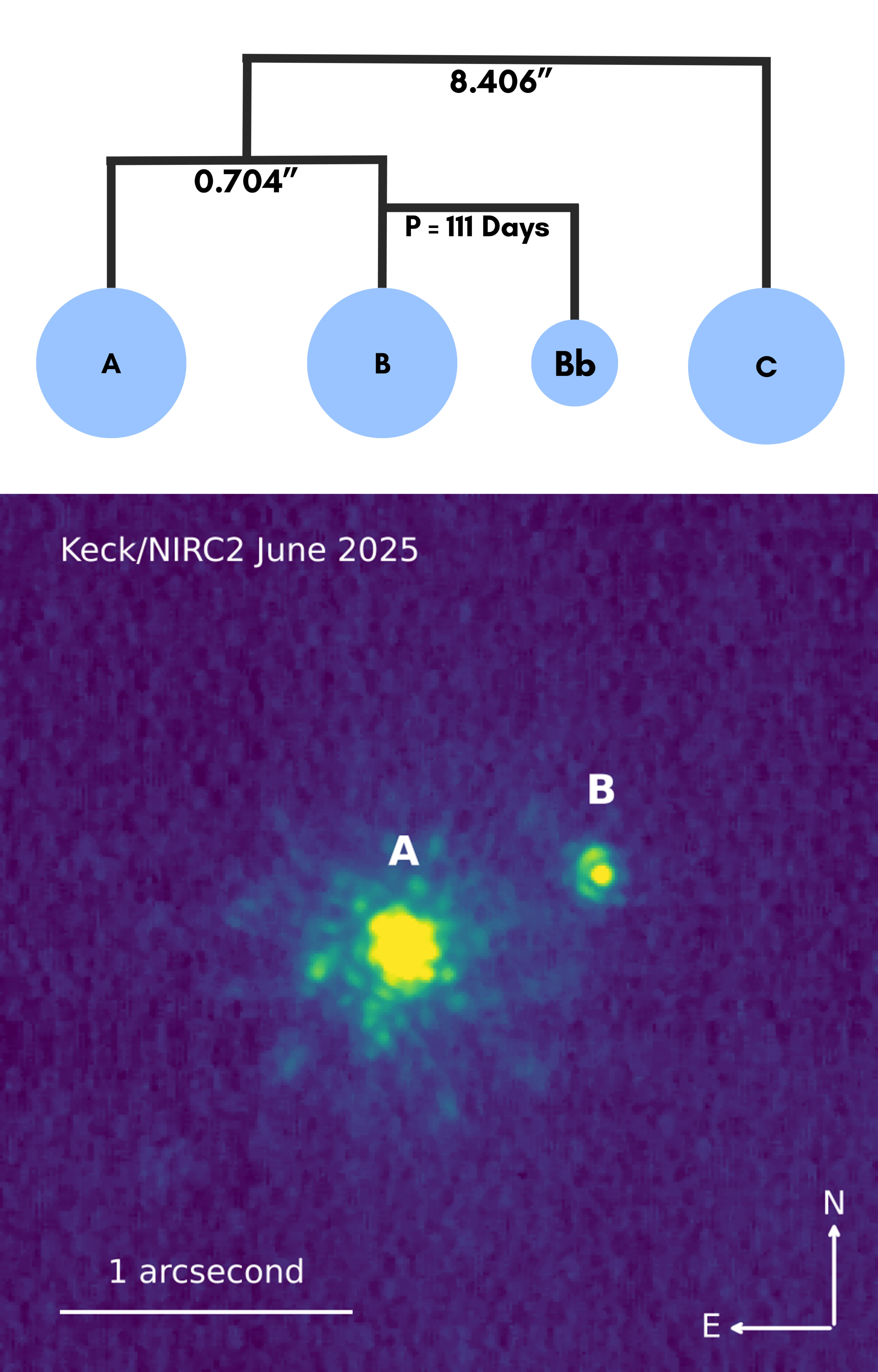}
  \caption{Top: Schematic of the hierarchical configuration of the KOI-1623 system (not to scale). Bottom: Keck/NIRC2 imaging of KOI-1623AB obtained on June 11 2025 with the $K^{\prime}$ filter, displayed using an asinh stretch. The field of view is approximately $3\arcsec \times 3\arcsec$.}
\label{fig:image}
\end{figure}

We obtained Keck/NIRC2 observations of the KOI-1623 system on 2025 June 11 \textsc{ut}. The target had previously been observed on 2012 May 07 \textsc{ut} (PI Crepp) using the $J$ and $K^{\prime}$ filters, and on 2013 June 15 \textsc{ut} (PI Marcy) using the $K^{\prime}$ filter. These earlier epochs clearly revealed a binary companion (KOI-1623B). Our 2025 observations, obtained with the $K^{\prime}$ filter (PI Huber), also include non-redundant aperture masking (NRM). Figure~\ref{fig:image} shows our 2025 Keck/NIRC2 observation of KOI-1623. The imaging observations were reduced following the procedures described in \citet{2016AJ....152....8K}. We analyze the NRM data following the procedures described in \citet{Ireland2008ApJ...678..463I} and \citet{2016AJ....152....8K}, using the Sydney pipeline.\footnote{https://github.com/mikeireland/idlnrm} No additional companions beyond the known binary are detected. The 99.9\% confidence NRM detection limits, combined with those from the May 2012 observation, are presented in Table~\ref{tbl:contrast}, along with the contrast limits for the secondary star. From the 2025 epoch analysis, we derive a separation of $703.75 \pm 1.52$\,mas, a position angle of $289.999^\circ \pm 0.117^\circ$, and a flux contrast of $\Delta K^{\prime} = 2.799 \pm 0.008$\,mag, along with the corresponding contrast curves. The 2MASS catalog reports a $K_S$-band magnitude of $10.481\pm0.022$\,mag, encompassing flux from both stars \citep{Skrutskie2006AJ....131.1163S}. Using the measured magnitude difference $\Delta K^{\prime}=2.799 \pm 0.008$\,mag, we derive $K=10.559\pm0.022$\,mag for the primary.  

\begin{table*}[ht] 
\centering
\caption{Contrast limits in the $K$-band as a function of angular separation for KOI-1623AB.}
\renewcommand{\arraystretch}{1.4}
\begin{tabular}{ll*{16}{c}}
\hline\hline
\multicolumn{2}{c}{} & \multicolumn{16}{c}{Separation (mas)} \\
\cline{3-18}
Star & & 15 & 30 & 50 & 60 & 100 & 120 & 150 & 200 & 250 & 300 & 400 & 500 & 700 & 1000 & 1500 & 2000 \\
\hline
KOI-1623\,A & $\Delta K^{\prime}$ (mag) & $-0.5$ & 2.3 & \nodata & 3.2 & \nodata & 3.1 & 4.8 & 5.8 & 6.3 & 6.7 & 7.0 & 7.3 & 7.4 & 7.9 & 7.8 & 8.1 \\
KOI-1623\,B & $\Delta K^{\prime}$ (mag) & \nodata & \nodata & 0.0 & \nodata & 2.0 & \nodata & 4.1 & 4.5 & 4.5 & 4.5 & 4.7 & 4.8 & 1.4 & 3.8 & 5.1 & 5.1 \\
\hline
\end{tabular}\label{tbl:contrast}
\flushleft Note: Contrast limits for the primary star at separations of 15--120~mas are derived from the NRM observations, while limits at 150--2000~mas are obtained from direct imaging. No NRM analysis is available for the companion. We therefore estimated the companion's contrast limits at small separations by assuming $\Delta m = 0.0$ at 50~mas based on visual inspection of the images, and linearly interpolating between this value and the measured contrast limit of $\Delta m = 4.1$ at 150~mas, yielding an adopted contrast limit of $\Delta m = 2.0$ at 100~mas. 
\end{table*}

Given the system's projected separation and \textit{Gaia} parallax of $2.31\pm0.02$\,mas, the two components are separated by $305\pm3$\,AU, corresponding to an orbital period of $\sim$ 4300 years for a circular orbit. \citet{2016AJ....152....8K} studied the field contamination rate in the \textit{Kepler} field and found negligible contamination for separations $\rho < 1500$ AU and mass ratios $q\,(M_{\mathrm{sec}}/M_{\mathrm{pri}}) > 0.4$. KOI-1623AB has $\rho \approx 310$ AU and $q \approx 0.55$, indicating that the secondary is likely gravitationally bound to the primary. Using Table~4 of \citet{Pecaut2013ApJS..208....9P}, the \texttt{isoclassify}-derived effective temperatures from Section~\ref{sec:star} ($T_{\mathrm{eff,pri}} = 5699\pm100$, $T{_\mathrm{eff,sec}} = 3909^{+44.8}_{-41.1}$) predict $\Delta(J-K) \approx 0.45$\,mag, consistent with the NIRC2 observations from May 2012 ($\Delta(J-K) = 0.43 \pm 0.01$\,mag), supporting that the companions are physically bound. Additionally, the companion's position in the 2025 NIRC2 image is consistent with that in 2012 within astrometric uncertainties. Given the primary star’s \textit{Gaia} proper motions
($\mu_{\alpha,\mathrm{primary}} = -1.35 \pm 0.02\,\mathrm{mas\,yr^{-1}}$,
$\mu_{\delta,\mathrm{primary}} = -18.69 \pm 0.02\,\mathrm{mas\,yr^{-1}}$),
a background star would have moved noticeably relative to the primary.

\subsubsection{High Resolution Spectroscopy with Keck/HIRES}\label{sec:spec}
A spectrum of KOI-1623A was obtained on 2012 Sep 24 \textsc{ut} using the High Resolution Echelle Spectrometer (HIRES; \citealp{1994SPIE.2198..362V}) on the 10-m Keck~I telescope as part of the California-\textit{Kepler} Survey \citep{2017AJ....154..107P, 2017AJ....154..108J}. Observations were taken with the C2 decker resulting in a resolving power of $\approx$\,50\,000. The exposure time was $\approx$ 8 minutes, resulting in a SNR of 70 per pixel at 550\,nm for each spectrum. The spectrum was reduced using standard procedures described in \citet{2010ApJ...721.1467H}. 

Inspection of the spectrum following \citet{2015AJ....149...18K} yielded no evidence for secondary lines, ruling out companions as faint as 1\% of the primary star’s flux. Assuming the primary is a G-type star, this corresponds to companions as faint as an M dwarf. However, detection is limited by the slit size. Companions with separations greater than $0.43\arcsec$ to $1.5\arcsec$ may not be detected, and the companion’s separation of $703.75 \pm 1.52$ mas may lie outside this range depending on the slit orientation. 

To derive atmospheric parameters of the primary star from the spectrum we used Specmatch-synth \citep{2015PhDT........82P}, which fits synthetic stellar spectra from \citet{2005A&A...443..735C}. Specmatch-synth has been extensively used for the characterization of exoplanet host stars, including for the California-\textit{Kepler} Survey, and has been extensively tested against other methods \citep{2017AJ....154..107P, 2018ApJ...861..149F}. The resulting atmospheric parameters are $T_{\mathrm{eff}}$ = $5687 \pm 100$\,K, [Fe/H] = $0.03 \pm 0.06$\,dex, log(g) = $4.0 \pm 0.1$\,dex and $v\sin{i}=2\pm1$\,km/s. The primary star is thus a slightly evolved Sun-like star with near-solar metallicity.

\subsubsection{\textit{Gaia} Detection of a Bound Tertiary Companion}\label{sec:triple}
\textit{Gaia} identified a nearby source (\textit{Gaia} DR3 2132051937879983744; hereafter KOI-1623C) at $\sim 8\farcs4$ from KOI-1623A with a $G$-band magnitude of $G = 19.8$, consistent with an M-dwarf companion. The source exhibits common proper motion with the primary star, indicating that it is likely a bound tertiary companion:
$\mu_{\alpha,\mathrm{tertiary}} = -1.61 \pm 0.38\,\mathrm{mas\,yr^{-1}}$,
$\mu_{\delta,\mathrm{tertiary}} = -18.22 \pm 0.42\,\mathrm{mas\,yr^{-1}}$. The top panel of Figure~\ref{fig:image} schematically illustrates the hierarchical triple configuration of the KOI-1623 system.

\begin{figure}
  \includegraphics[width=0.48\textwidth]{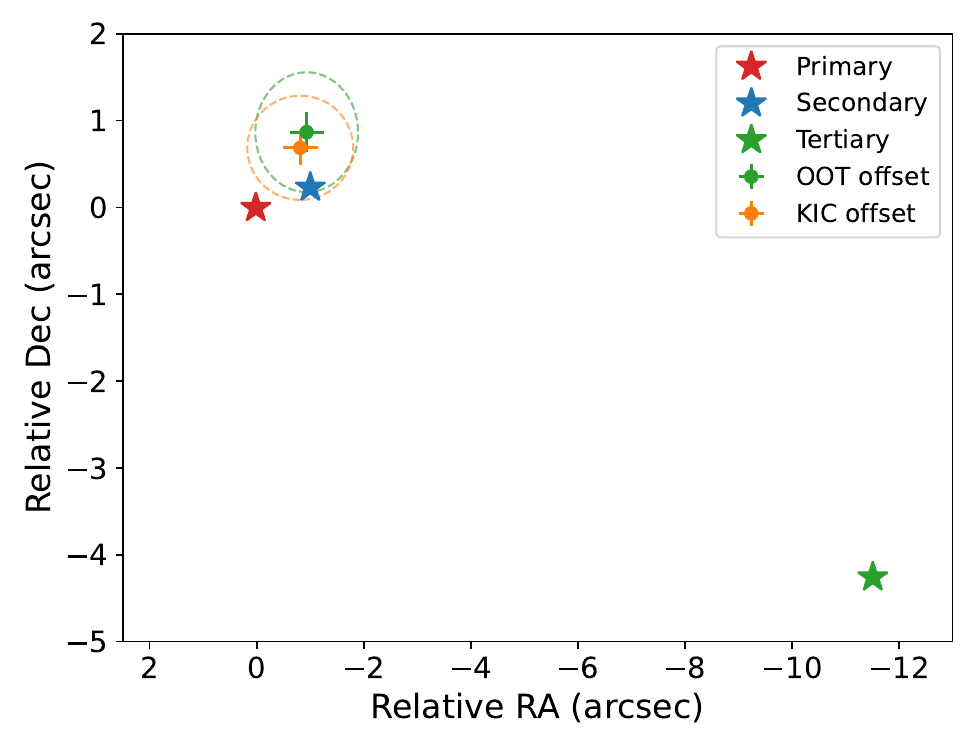}
  \caption{Sky positions of the primary (red), secondary (blue), and tertiary (green) stars in the KOI-1623 system. Solid circles with crosses indicate the Out-of-Transit (OOT) and KIC centroid offsets from \textit{Kepler}'s Data Validation Report, and dashed circles representing 3-sigma uncertainties.}
\label{fig:centroid}
\end{figure}

Figure~\ref{fig:centroid} shows all three star’s positions together with the centroid offsets with respect to the binary's flux-weighted photocenter. The centroid offsets are approximately aligned with the position angle connecting the primary and secondary stars. The secondary also lies within the $3\sigma$ uncertainty region of the offsets, suggesting preliminary evidence that the transit signal may originate from the secondary. The tertiary is well separated from the primary, secondary, and the centroid offset position, and is therefore unlikely to be the source of the transit signal. In the remainder of the paper, we focus on analyzing the inner binary.

\subsection{Stellar Parameters}\label{sec:star}
\begin{figure}
  \includegraphics[width=0.48\textwidth]{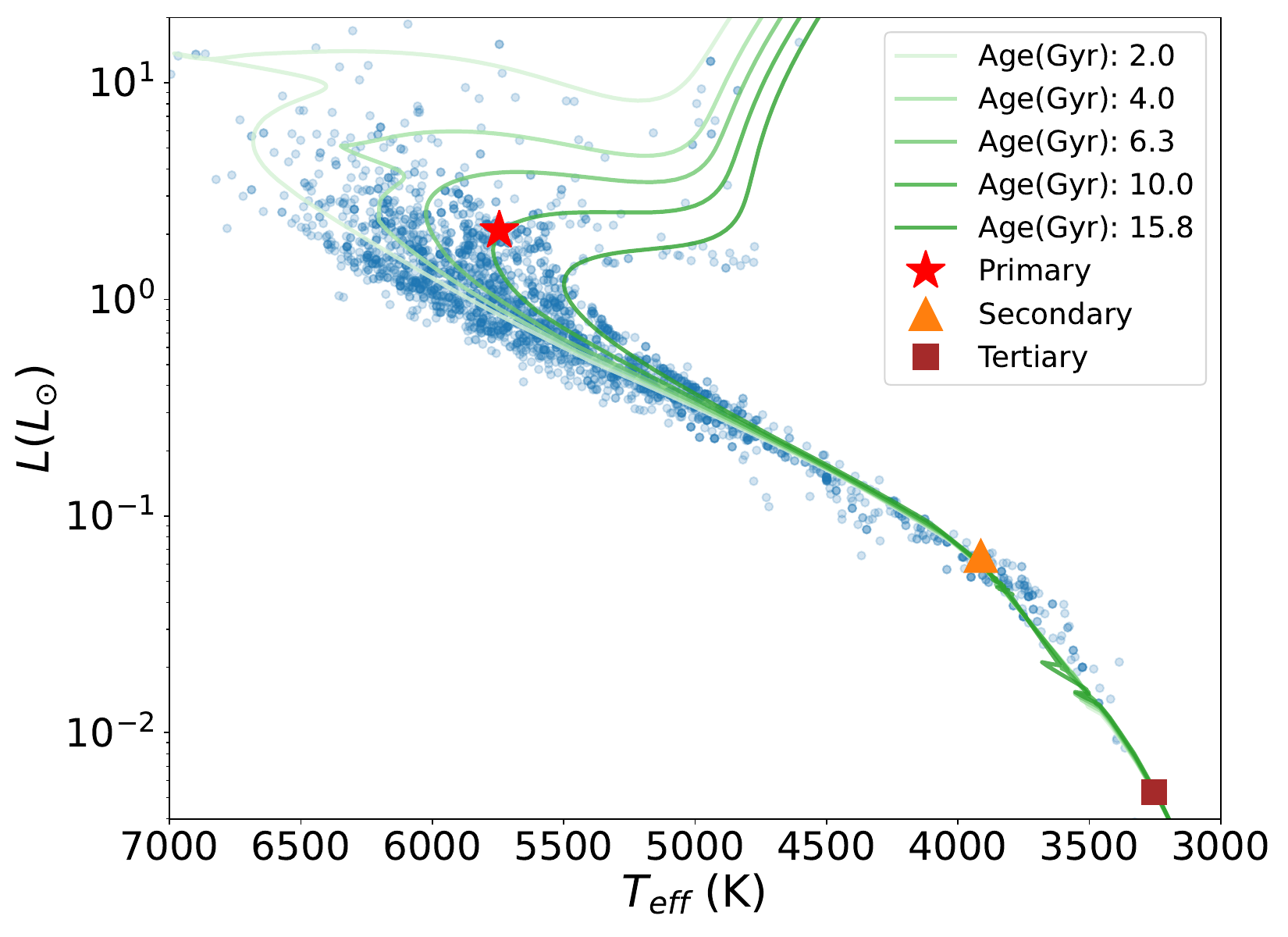}
  \caption{The stars of the KOI-1623 system on an H-R diagram, along with other \textit{Kepler} planet host stars. The stellar parameters are adopted from \citet{Berger2020AJ....159..280B}. The primary is shown as the red star, the secondary is shown as the orange triangle, and the tertiary is shown as the red square. Several isochrones are overplotted for reference.}
\label{fig:HR}
\end{figure}

We used the stellar classification code \texttt{isoclassify} to derive the parameters of the inner binary \citep{Huber2017ApJ...844..102H, Berger2020AJ....159..280B, Berger2023arXiv230111338B}. We adopt the MESA model grids \citep{Choi2016ApJ...823..102C} and used effective temperature, metallicity, $K$-band photometry, and parallax as input parameters. When provided with the contrast ratio of a secondary star, \texttt{isoclassify} constrains their relative luminosity difference, determines the companion’s position on the isochrone, and derives its stellar parameters under the assumption of coevality.

We find that the primary is a $1.022^{+0.043}_{-0.038} \,M_\odot$, $1.477\pm0.031 \,R_\odot$ G-type star, and the secondary is a $0.557^{+0.024}_{-0.021} \,M_\odot$, $0.564^{+0.012}_{-0.012} \,R_\odot$ star at the K/M boundary. We adopt the same fractional uncertainties for the secondary’s mass and radius as those of the primary. Figure~\ref{fig:HR} compares all three stars on an HR diagram with other \textit{Kepler} host stars with MIST isochrones overplotted \citep{Paxton2011ApJS..192....3P, Paxton2013ApJS..208....4P, Dotter2016ApJS..222....8D, Choi2016ApJ...823..102C, Paxton2018ApJS..234...34P}. The primary is slightly evolved off the main sequence and the secondary is a K-dwarf on the lower main sequence.  We also derive the stellar parameters of the tertiary using its 2MASS $K_S$-band magnitude of $15.447$\,mag \citep{Skrutskie2006AJ....131.1163S}. 2MASS does not report an uncertainty for this source, so we adopt an uncertainty of $\sim 0.2$\,mag, consistent with the typical precision at this magnitude \citep{2MASSCutri2003}. The resulting stellar parameters for the triple system are summarized in Table~\ref{tbl:stellar_params}.

\begin{table}[ht]
\centering
\caption{Derived Stellar Parameters for the Triple-Star System KOI-1623.}
\label{tbl:stellar_params}
\begin{tabular}{lccc}
\hline
\multicolumn{4}{l}{\textbf{Isoclassify Inputs}} \\
\hline
$T_{\rm eff}$ (K)         & \multicolumn{2}{c}{$5687 \pm 100$\phn} \\
{[Fe/H]}                  & \multicolumn{2}{c}{$0.03 \pm 0.06$} \\
Parallax (\arcsec)     & \multicolumn{2}{c}{$0.0023113 \pm 0.0000174$} \\
$K_S$ (mag)          & \multicolumn{2}{c}{$10.559 \pm 0.022$\phn} \\
$\Delta K_{B-A}$ (mag)      & \multicolumn{2}{c}{$2.799 \pm 0.008$} \\
$\Delta K_{C-A}$ (mag)      & \multicolumn{2}{c}{$4.887 \pm 0.200$} \\
\hline
\multicolumn{4}{l}{\textbf{Derived Parameters}} \\
Parameter & Primary & Secondary & Tertiary \\
\hline
Mass ($M_\odot$)   & $1.022^{+0.043}_{-0.038}$ & $0.557^{+0.024}_{-0.021}$&$0.197^{+0.180}_{-0.051}$ \\
Radius ($R_\odot$)    & $1.477^{+0.031}_{-0.031}$ & $0.564^{+0.012}_{-0.012}$ &$0.232^{+0.160}_{-0.052}$\\
Age (Gyr)                 & $9.06^{+1.45}_{-1.50}$    & --                  & -- \\
$\rho$ ($\rho_\odot$) & $0.31^{+0.03}_{-0.03}$ & $3.09^{+0.09}_{-0.08}$ &$13.72^{+7.06}_{-8.27}$\\
$T_{\rm eff}$ (K)      & $5699^{+101}_{-99}$ & $3909^{+45}_{-41}$ & $3253^{+347}_{-96}$\\
$\log g$ (cgs)         & $4.106^{+0.030}_{-0.028}$ & $4.680^{+0.006}_{-0.007}$& $4.975^{+0.083}_{-0.169}$\\
\hline
\end{tabular}
\flushleft Note: Quoted uncertainties are random errors and do not include systematic errors due to different input physics in model grids \citep{Tayar2022ApJ...927...31T}.
\end{table}

\subsection{Transit Fit} \label{sec:transit}

\begin{figure*}
  \includegraphics[width=\textwidth]{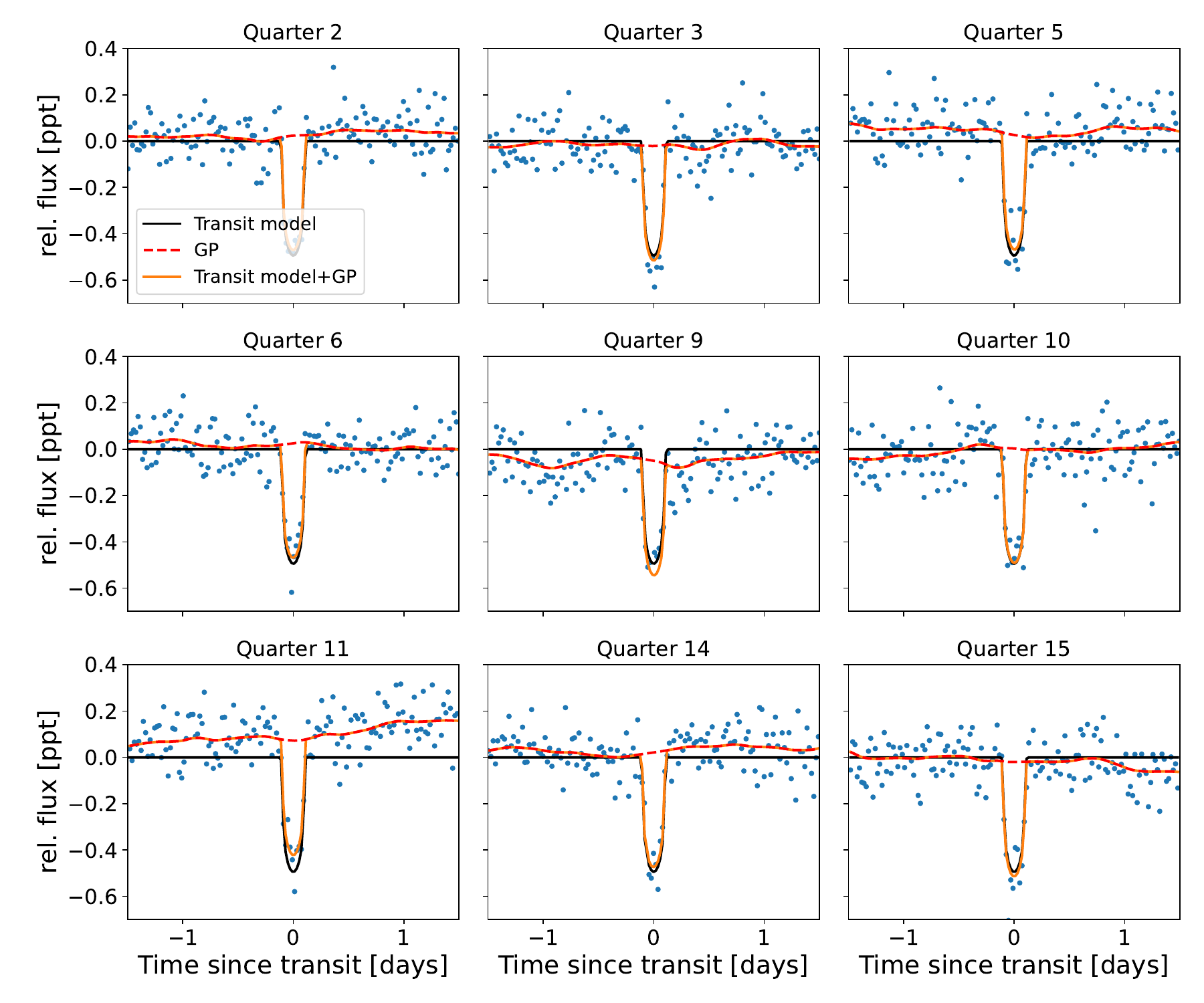}
  \caption{Transit fit for KOI-1623.01 across all available quarters with clear transit detections. Blue points: observed light curve data. Black curve: best-fit transit model. Red curve: Gaussian Process (GP) model accounting for stellar variability and systematics. Orange curve: combined transit + GP model.}
\label{fig:transit}
\end{figure*}

\begin{figure*}
  \includegraphics[width=\textwidth]{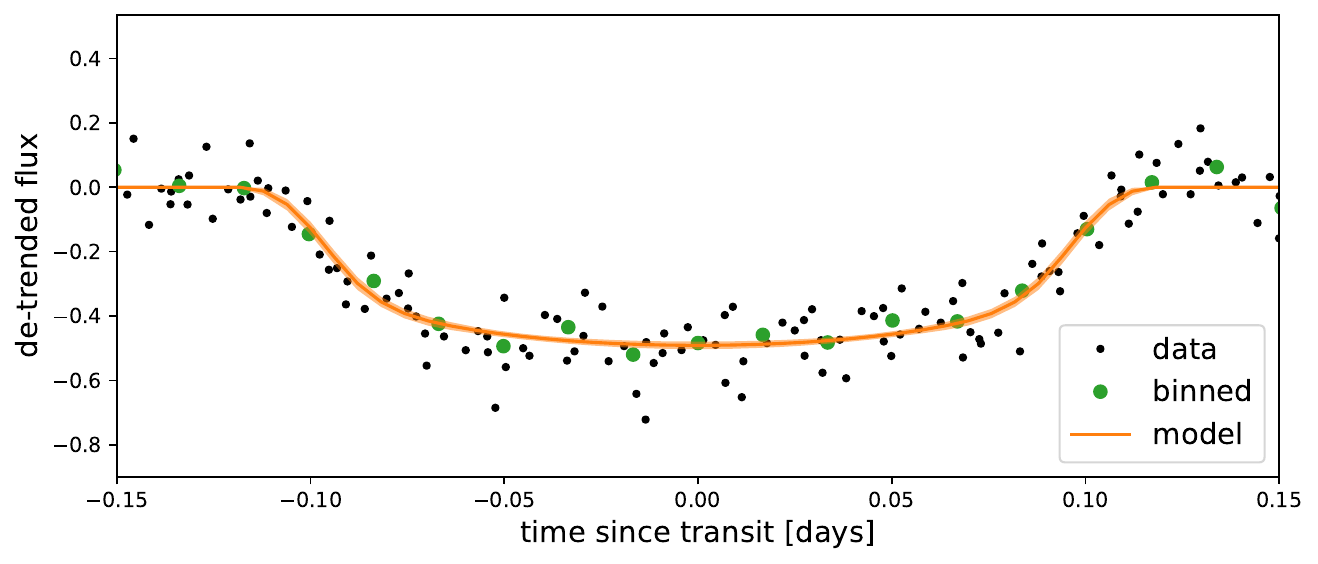}
  \caption{Black points: Phase-folded light curve of KOI-1623.01 after removing stellar variability using a Gaussian Process model. Green points: Binned light curve. Orange line: Best-fit transit model with shaded uncertainty representing the 16th–84th percentile range.}
\label{fig:folded-transit}
\end{figure*}

\subsubsection{Planetary Parameters}
\begin{figure*}
  \includegraphics[width=\textwidth]{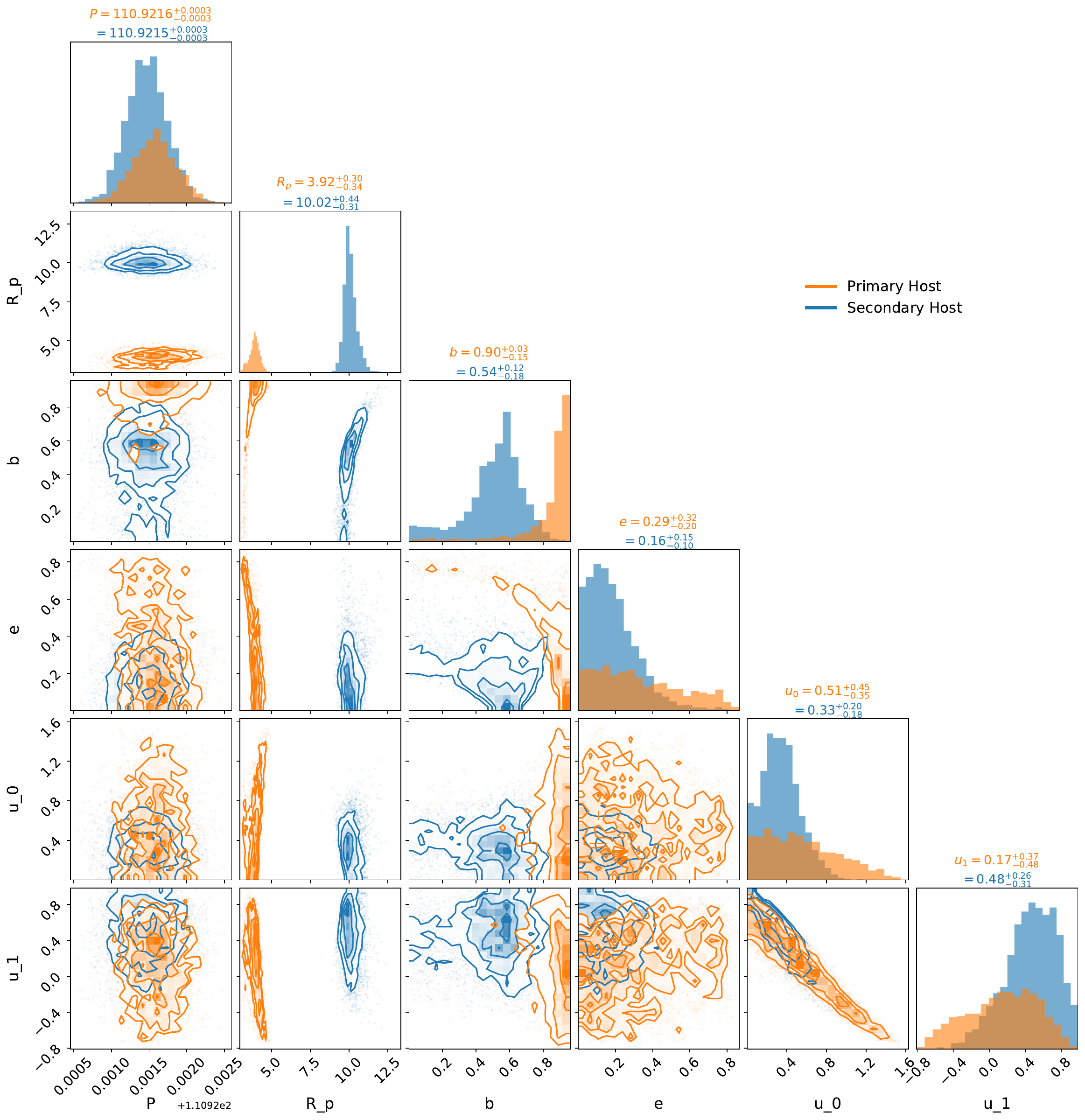}
  \caption{Corner plot of the transit fit for KOI-1623.01, assuming the planet transits either the primary (orange) or secondary (blue) star.}
\label{fig:corner_compare}
\end{figure*}
We first derive the planetary parameters under the assumption that the planet is orbiting either the primary or the secondary separately. We denote these two hypotheses KOI-1623Ab (the planet orbiting the primary) and KOI-1623Bb (the planet orbiting the secondary), and carry both through the analysis below. 

We fit the transit light curve using the \texttt{exoplanet} package \citep{exo2021JOSS....6.3285F} with priors on the stellar densities for each component. We adopt non-informative priors on the limb-darkening coefficients, following \citet{Kipping2013MNRAS.435.2152K}. We also specify stellar density by adopting bounded normal priors on the stellar mass and radius, using \texttt{isoclassify}-derived means and uncertainties.

We perform a Fourier analysis of the light curve and identify a characteristic $1/f$-like profile, consistent with stellar granulation \citep[e.g.,][]{Harvey1985ESASP.235..199H,Kallinger2014A&A...570A..41K}. To account for this signal, we model the light curve with Gaussian Process (GP), applying two, stochastically-driven, damped harmonic oscillator (SHO) kernels. For one kernel, we fix the quality factor to $Q = 1/{\sqrt{2}}$, while allowing $Q$ to vary for the other kernel to capture additional variability components. 
Figure~\ref{fig:transit} shows the transit model jointly fit with a Gaussian Process to each quarter of data individually, capturing both the planetary signal and correlated noise. Figure~\ref{fig:folded-transit} shows the phase-folded light curve after removing the Gaussian Process stellar variability model, along with the best-fit transit model.

Flux dilution in binary systems alters the observed transit depth, making transits appear shallower due to additional flux from the companion star, and thus requires corrections to the derived planetary radius \citep{Furlan2017AJ....153...71F}. We convert the $K$-band magnitude into the \textit{Kepler} band by interpolating the MIST isochrone with the isoclassify outputs - mass, age, metallicity, distance, and extinction \citep{morton2015ascl.soft03010M}. We derive a \textit{Kepler}-band contrast of $\Delta Kp = 4.5$\,mag and use it to correct the planet radius following Equations (3)-(7) of \citet{Furlan2017AJ....153...71F}. 

Figure~\ref{fig:corner_compare} presents the posterior distributions from the transit fit, assuming the planet orbits either the primary or secondary component of the binary system. The inferred planet size is highly sensitive to which star it transits: if orbiting the primary, the planet would be Neptune-sized with a radius of $3.92^{+0.30}_{-0.34}\,R_\Earth$, whereas a transit across the secondary would imply a giant planet with a radius of $10.02^{+0.44}_{-0.31}\,R_\Earth$. The main fitted planetary parameters are summarized in Table~\ref{tbl:param}.

\begin{table}[ht]\label{tbl:stellar params}
\centering
\caption{Planet Parameters Assuming Primary vs. Secondary Host}
\label{tbl:param}
\begin{tabular}{lcc}
\hline
Parameter & Primary Host & Secondary Host\\
\hline
Period (days)   & \multicolumn{2}{c}{$110.9216^{+0.0003}_{-0.0003}$} \\ 
Planet Radius ($R_\Earth$)   &$3.92^{+0.30}_{-0.34}$ & $10.02^{+0.44}_{-0.31}$ \\
Impact Parameter &$0.90^{+0.03}_{-0.15}$ &$0.54^{+0.12}_{-0.18}$\\
Eccentricity &$0.29^{+0.32}_{-0.20}$ &$0.16^{+0.15}_{-0.10}$\\
\hline
\end{tabular}
\end{table}

\subsubsection{Stellar Inferred Density}
\begin{figure}

  \includegraphics[width=0.50\textwidth]{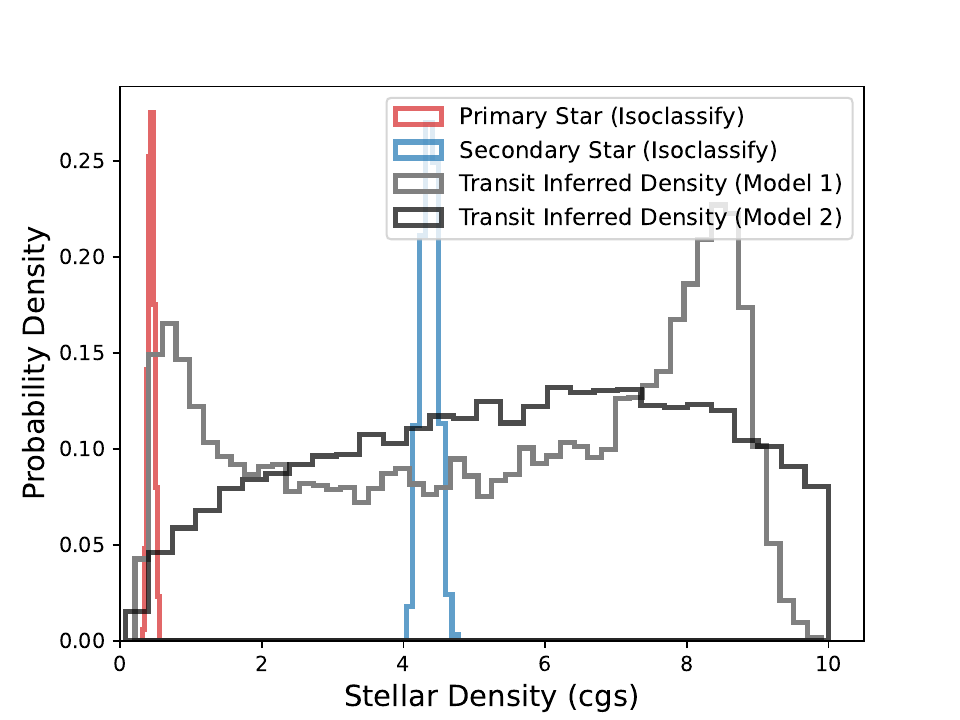}
  \caption{Posterior distribution of the planet host star's inferred density from the transit fit using the \texttt{ALDERAAN} model (gray, Model 1) and the \texttt{pymc} model (dark gray, Model 2), overlapped with stellar density distributions of the primary star (red) and the secondary star (blue), derived from \texttt{isoclassify}. The inferred density is shown as a probability density. For the isoclassify density distributions, which are much narrower, we instead plot raw counts with uniform weights scaled by $\frac{1}{1000}$ to allow better visual comparison.}
\label{fig:density}
\end{figure}

We first attempt to identify the star responsible for the transit by deriving the host star's stellar parameters from the transit fit alone, without imposing any informative priors on the stellar density. Four key parameters can be derived from transit observables (transit depth, duration, and period): the planet-to-star radius ratio $\frac{R_{p}}{R_{*}}$, the impact parameter $b$, the ratio $\frac{a}{R_{*}}$, and the stellar density $\rho_*$. Specifically, stellar density can be determined by combining $\frac{a}{R_{*}}$ with Kepler's third law, with assumptions such as a circular orbit and $M_p\ll M_*$ \citep{Seager2003ApJ...585.1038S}. 

A challenge in transit modeling is that the stellar density is strongly correlated with the impact parameter, $b$, the planet-to-star radius ratio $\frac{R_p}{R_*}$, and the transit duration. This degeneracy implies that a planet could either transit a larger, lower-density star with a higher impact parameter or a smaller, higher-density star with a lower impact parameter. Consequently, constraining the impact parameter more precisely is critical. This issue becomes particularly significant for grazing and near-grazing transits, where the correlation between the impact parameter and the planet-to-star radius ratio is especially strong \citep{umbrella_transit}. 

To address this, we apply umbrella sampling to better capture the geometry of the orbit. Originally introduced by \citet{umbrella}, umbrella sampling was only recently applied to transit fitting by \citet{umbrella_transit}. Umbrella sampling divides a target distribution into overlapping windows, applies bias functions to flatten local peaks and valleys, and then recombine the samples into a complete posterior distribution. This method is designed to more effectively sample from multi-modal distributions; in our case, it enables better exploration of the grazing transit regime. We use the transit fitting code \texttt{ALDERAAN}\footnote{https://github.com/gjgilbert/alderaan}, which detrends the light curves using a Gaussian Process model with a rotation kernel and applies umbrella sampling to more thoroughly explore the parameter space \citep{umbrella_transit}. \texttt{ALDERAAN} requires photometric light curves, which we obtain from the Mikulski Archive for Space Telescopes (MAST).

We also construct a second transit model that allows for orbital eccentricity and uses a standard sampling approach. In this model, we used \texttt{lightkurve} to access and normalize the \textit{Kepler} light curves and removed systematics using the default \texttt{lightkurve} functions and parameters. We use \texttt{exoplanet} and \texttt{pymc} to generate and sample transit models, using routines that follow \citet{Burns-Watson2026AJ....171..327B}. However, compared to \citet{Burns-Watson2026AJ....171..327B}, this analysis uses version 5 of \texttt{pymc} instead of \texttt{pymc3} and includes eccentricity and argument of periastron as free parameters within the model. In addition to these parameters, the model includes priors for common transit fitting parameters such as orbital period, $R_p/R_\star$, impact parameter, stellar density, and quadratic limb darkening. The results reported here come from 4 chains that each ran 6000 draws after running 8000 tuning steps.

Figure~\ref{fig:density} compares the transit-inferred stellar density distributions from both models to the isochrone-derived densities. The two models produce different posterior shapes, primarily due to their treatment of the impact parameter and orbital eccentricity. 

In particular, since grazing transits are correlated with lower inferred stellar densities, the weighting of the grazing transit regime in the \texttt{ALDERAAN} sampling may be responsible for the low density peak. The \texttt{ALDERAAN} results also used a model that assumed a circular orbit, whereas the \texttt{pymc} fit treated eccentricity as a free parameter. This could also contribute to the difference the posteriors, especially between the peak values of the posteriors ($\sim7 g/cm^3$ from the \texttt{pymc} sampling versus $\sim9 g/cm^3$ from the \texttt{ALDERAAN} sampling). \citet{Kipping2014MNRAS.440.2164K} notes that when assuming circular orbits, the stellar density is likely to be overestimated if the true eccentricity is non-trivial. More broadly, allowing for eccentric orbits allows for a wider range of inferred stellar densities. In this case, the transit-inferred density alone is insufficient to confidently identify the planet’s host star, as the distribution can reasonably correspond to the densities of either stellar component.

Using the posterior distributions in Figure \ref{fig:density}, we calculate circumsecondary probabilities of 37\% for the \texttt{ALDERAAN} model and 74\% for the \texttt{pymc} model. We conclude that the transit-inferred density alone is insufficient to confidently identify the planet’s host star, as the distribution can reasonably correspond to the densities of either stellar component.
\subsection{Identifying The Planet Host with Pixel Level Analysis}\label{sec:pixel}

So far, our analysis has relied on the \textit{Kepler} out-of-transit (OOT) and KIC centroid offsets, along with the transit light curve. Although Figure~\ref{fig:centroid} strongly suggests that the secondary is the source of the transit signal, it does not definitively identify the planet-hosting star. To further constrain the transit source, we use the \texttt{Transit-APP} tool \citep{Bryson_2025}, which analyzes pixel-level observations to calculate the probability that the transit originates from either the primary star, the companion star, or another nearby source.

\texttt{Transit-APP} uses a series of difference images, constructed by subtracting out-of-transit pixel images from in-transit pixel images. These difference images localize the transit signal on the detector and are compared with simulated difference images generated for the target star and all nearby stars. The agreement between the observed and simulated images is then used to calculate the probability that each star is the source of the transit signal.

\begin{figure*}
  \includegraphics[width=\textwidth]{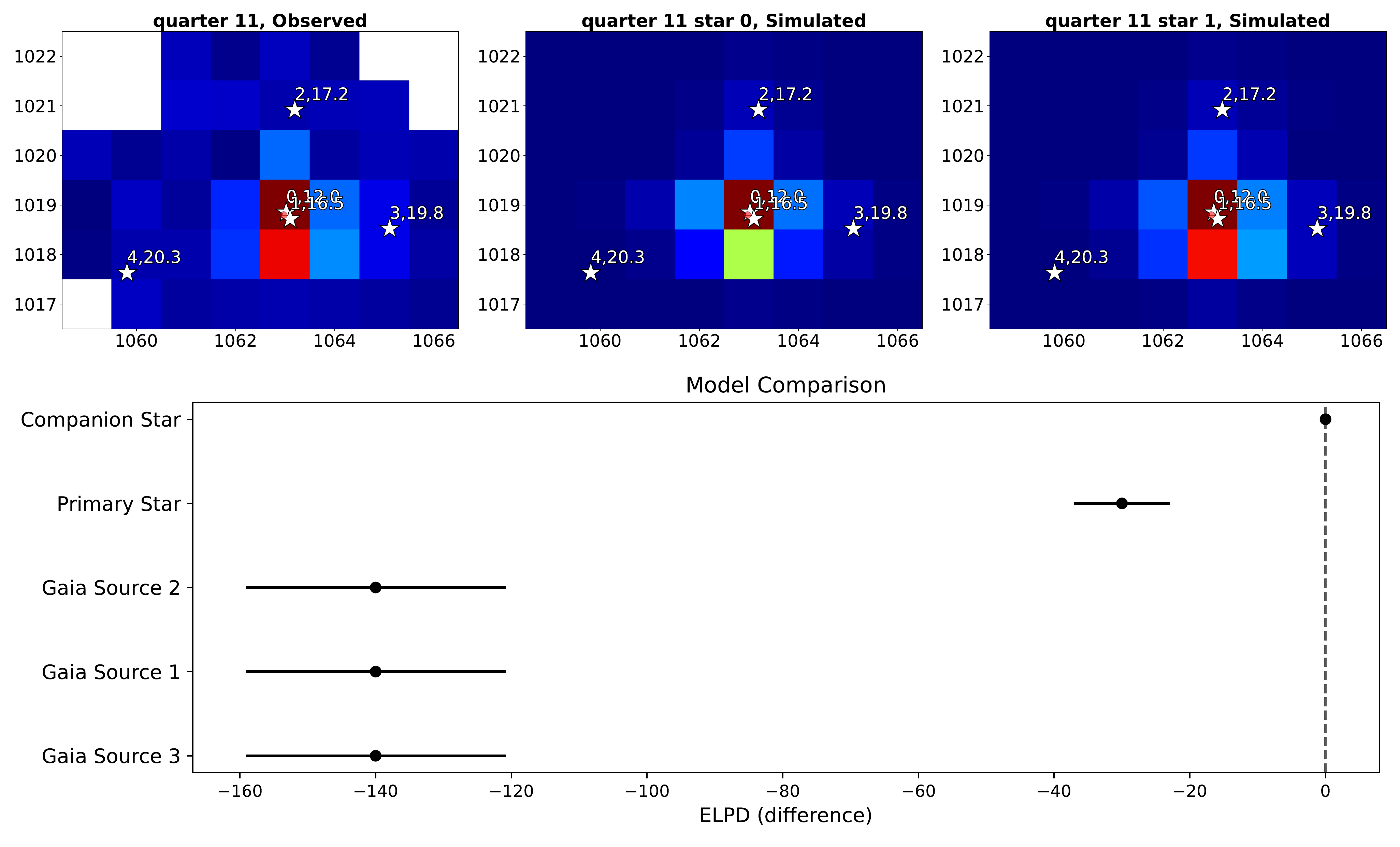}
  \caption{Top panel: Observed and simulated transit difference images for KOI-1623.01 in Quarter 11. Stars 0 and 1 correspond to the primary and companion stars, respectively. Stars 2, 3, and 4 are nearby Gaia sources in the field, including the tertiary companion (star 3). The Gaia magnitudes of all sources are labeled next to the corresponding stars. The simulated difference image for the companion star provides a significantly better match to the observed transit signal, supporting the conclusion that the companion is the source of the transit. Bottom panel: Model comparison with the expected log pointwise predictive density (ELPD) metric for all stars in the image, where the best-fitting model is normalized to $\Delta \mathrm{ELPD}=0$. The companion is strongly favored to be the transit host.
}
\label{fig:pixel}
\end{figure*}

We applied this pixel-level analysis to all quarters containing observed transits, with an example fit to the quarter~11 difference image shown in the top panel of Figure~\ref{fig:pixel}. Considering all quarters jointly, the analysis strongly favors the companion star as the source of the transit signal.
To compare models, we use the expected log pointwise predictive density (ELPD), a leave-one-out cross-validation metric that quantifies predictive accuracy on held-out data, with larger values indicating better predictive performance \citep{Vehtari2015arXiv150704544V}. For each candidate host, we calculate the ELPD difference relative to the best-fitting model ($\Delta$ELPD) and its standard error (dse), both provided by \texttt{Transit-APP}. The bottom panel of Figure~\ref{fig:pixel} shows the $\Delta$ELPD values for all sources in the field. The companion star is clearly preferred over the primary ($\Delta$ELPD $= 27.7 \pm 6.9$, i.e.\ $4.0\times$ its standard error) and is decisively favored over all nearby Gaia sources (best case $\Delta$ELPD $= 141.6 \pm 18.8$). Bayesian stacking \citep{Yao2017arXiv170402030Y} assigns the companion a weight $>0.999$, representing its contribution to optimal predictive performance when combining all models and can be interpreted as an approximation to the probability that the model is correct relative to the other candidate host stars. We therefore identify the companion star as the host of the transiting planet, rather than the primary or any other known nearby source.

\subsection{Statistical Validation Analysis with \texttt{TRICERATOPS} }\label{sec:validation}

Radial velocity (RV) measurements are usually required to confirm candidate planets. However, the faintness and large number of \textit{Kepler} planet hosts make such follow-up observations challenging \citep{Morton2016ApJ...822...86M}. Consequently, a significant fraction of \textit{Kepler}'s confirmed exoplanets were statistically validated by calculating their false positive probabilities without requiring additional follow-up observations. This method was first demonstrated for a small number of systems by \citet{Torres2011ApJ...727...24T} and later extended to a larger sample \citep[e.g.,][]{Morton2016ApJ...822...86M, triceratops}. 

Based on the pixel-level analysis presented in Section~\ref{sec:pixel}, which identifies the companion star as the transit host, we perform a validation analysis of KOI-1623Bb using \texttt{triceratops}, which uses a Bayesian framework to determine the likelihood that a transit signal originates from a true planet, while considering false positive scenarios and contamination from nearby stars. To reflect the results of the pixel-level analysis, we modify the \texttt{triceratops} input catalog by replacing the original target star (KOI-1623A) with the resolved companion (KOI-1623B) as the target, while retaining KOI-1623A as a nearby star. This allows \texttt{triceratops} to account for the flux contribution from KOI-1623A when modeling the transit light curve. We adopt the optimal aperture reported in the \textit{Kepler} Data Validation (DV) report for each quarter, which is used to identify the stars contributing significant flux within the aperture. We assign the companion the stellar properties derived in Section~\ref{sec:star} (mass, radius, and effective temperature), together with its astrometric information and the NIRC2 contrast curve (Table~\ref{tbl:contrast}). To satisfy the input requirements of \texttt{triceratops}, we estimate the companion's TESS magnitude by adding the \texttt{isoclassify}-derived $I$-band contrast, $\Delta I_{\mathrm{B-A}} = 4.01 \pm 0.09$ mag (Section~\ref{sec:star}), to the primary's TESS magnitude ($T = 11.55$). We then exclude scenarios in which the transit originates from other neighboring stars, including the primary (i.e., eclipsing binaries or transiting planets associated with nearby sources). The \texttt{triceratops} scenario definitions are listed in Table~\ref{tab:triceratops_scenarios}. 

\texttt{Triceratops} requires user-provided light curves as input. We use the \texttt{lightkurve} package to download the light curves for KOI-1623.01 \citep{lightkurve}. We apply the same data processing and modeling approach used in the transit fitting for consistency. The light curve was pre-cleaned using the same sigma-clipping procedure, and the stellar variability was modeled using a two-component SHO Gaussian Process. 

\begin{figure}

  \includegraphics[width=0.48\textwidth]{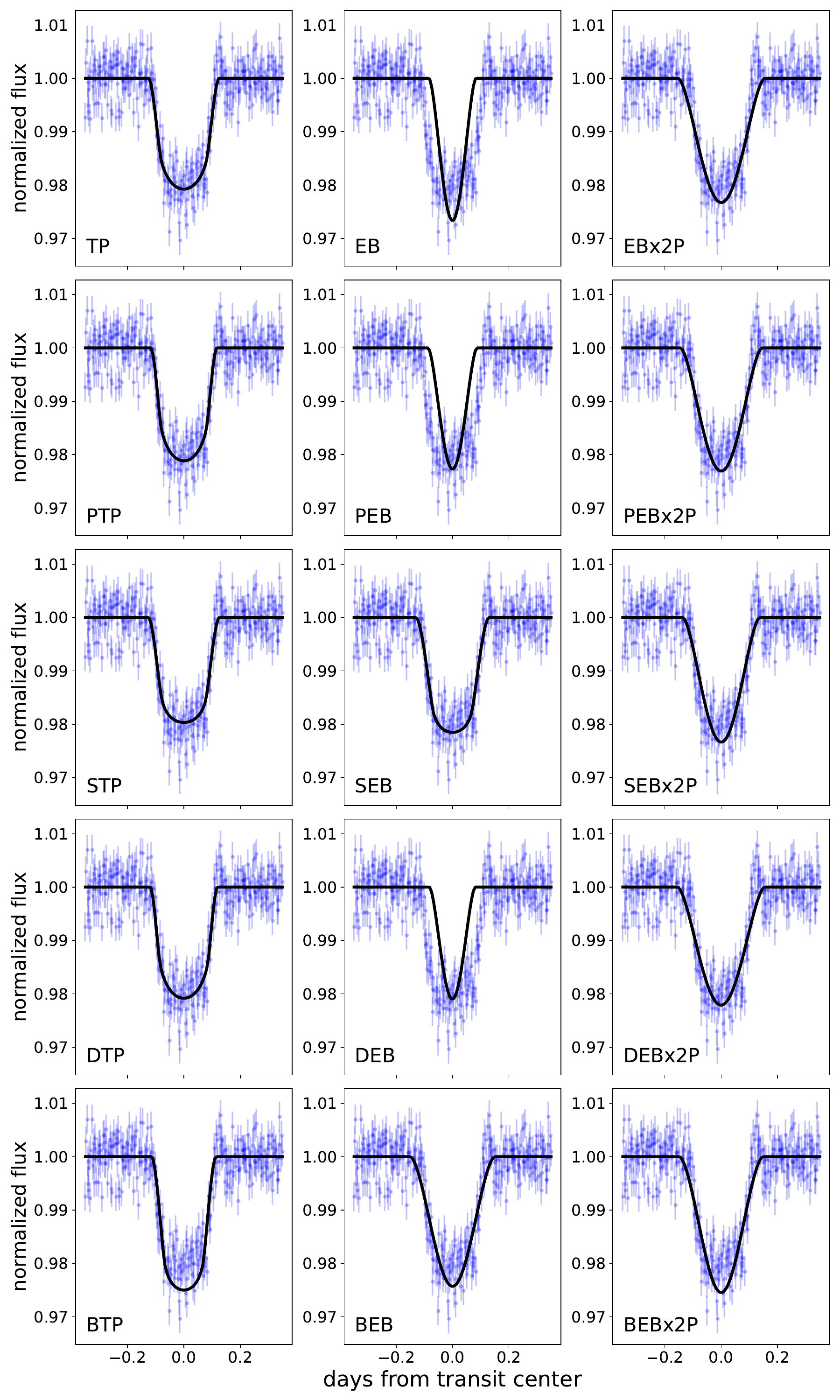}
  \caption{Transit fit of all transit scenarios for KOI-1623.01, assuming the companion star is the transit host. The scenarios being fit for is in the lower left corner of each panel. The most likely scenarios are as follows: TP — a transiting planet orbiting the companion star; DTP — a transiting planet orbiting the companion star with flux dilution from an unresolved background source; and PTP — a transiting planet orbiting the companion star with flux dilution from an unresolved bound companion.}
\label{fig:valid}
\end{figure}

\texttt{Triceratops} classifies candidates using the false positive probability (FPP) and nearby false positive probability (NFPP). Candidates are considered validated planets if $\mathrm{FPP}<0.015$ and $\mathrm{NFPP}<10^{-3}$, likely planets if $\mathrm{FPP}<0.5$ and $\mathrm{NFPP}<10^{-3}$, and likely nearby false positives if $\mathrm{NFPP}>10^{-1}$, following the criteria recommended by \citet{triceratops}. The FPP and NFPP are defined in Equations 4 and 5 of \citet{triceratops}. These criteria are broadly consistent with the exoplanet validation literature, where a false positive probability of FPP $<0.01$ is generally adopted as the validation criterion \citep{Morton2012ApJ...761....6M,Rowe2014ApJ...784...45R, Morton2016ApJ...822...86M}. Because our pixel-level analysis rules out nearby stars as potential transit hosts, we focus on the FPP in our validation analysis. We include all \texttt{triceratops} scenarios in the initial calculation, but subsequently assign zero probability to the nearby false-positive scenarios and renormalize the remaining probabilities before computing the final FPP.

We found that the calculated false positive probabilities vary between runs, likely due to the large orbital period introducing additional Poisson noise, as the code samples a wide range of orbital inclinations, including those that do not produce transits. To account for this, we ran the validation code 1000 times and computed the FPP and NFPP for each run. The fitting results for one of the runs is shown in Figure~\ref{fig:valid}, and the scenario definitions along with their mean probabilities across 1000 runs are listed in Table~\ref{tab:triceratops_scenarios}. Several degenerate scenarios are present, with the associated probabilities varying between runs. 

\begin{figure}
  \includegraphics[width=0.48\textwidth]{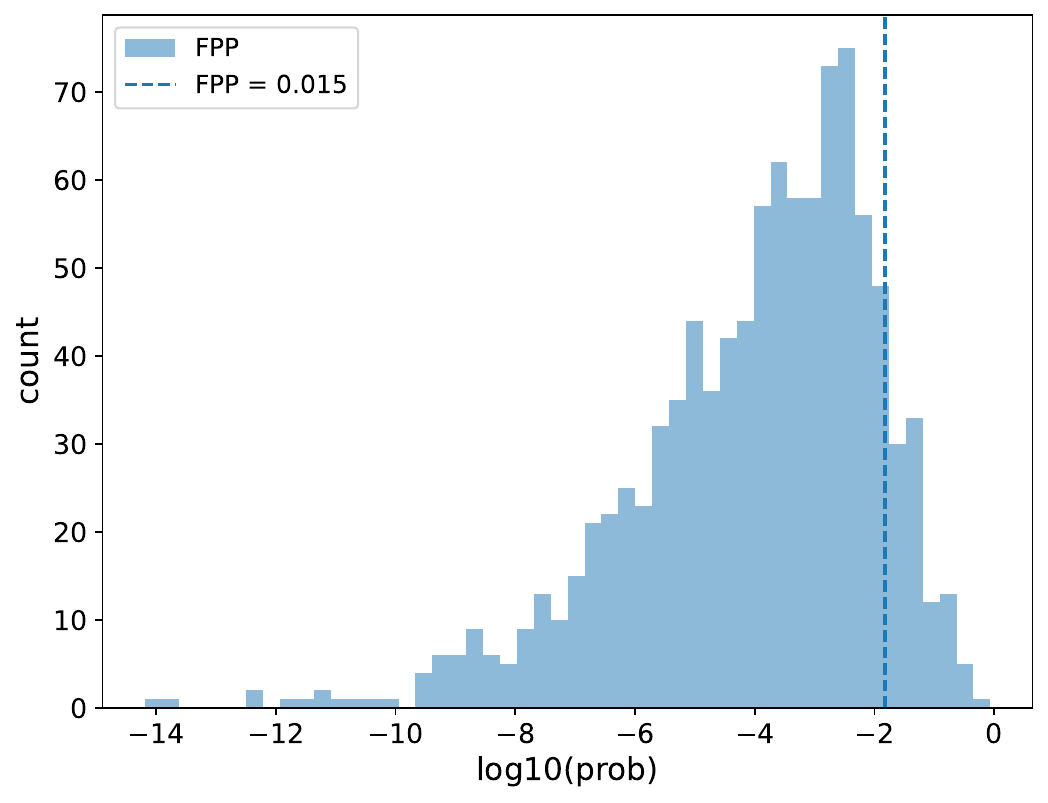}
  \caption{Distributions of the False Positive Probability (FPP) from 1000 runs of \texttt{triceratops}. The majority of the realizations yield FPP values below the adopted validation threshold.}
\label{fig:valid_stat}
\end{figure}

Figure~\ref{fig:valid_stat} presents the distribution of FPP and NFPP across these 1000 iterations. The median FPP and NFPP are $1.07 \times 10^{-6}$ and $1.81 \times 10^{-7}$, respectively, while the corresponding means are $0.0058$ and $2.14 \times 10^{-6}$. Both quantities lie below the statistical thresholds adopted by \citet{triceratops} ($\mathrm{FPP}<0.015$ and $\mathrm{NFPP}<10^{-3}$). Combined with our pixel-level analysis, which rules out neighboring stars as potential transit hosts, these results indicate that common false-positive scenarios considered by \texttt{triceratops} are strongly disfavored for KOI-1623.01. However, if the planet is transiting the companion star, it will be a giant planet with radius of $10.02^{+0.44}_{-0.31}\,R_\Earth$, and the planetary radius becomes degenerate with that of a low-mass star or brown dwarf, and the transit photometry alone cannot distinguish these cases from a true planet \citep{Mayo2018AJ....155..136M, triceratops}. We therefore do not claim formal planetary validation and instead interpret the \texttt{triceratops} results as demonstrating that the remaining significant false-positive scenario is a low-mass star or brown dwarf.

\section{KOI-1623.01 in Context with Exoplanet Demographics}

\begin{figure*}
  \includegraphics[width=\textwidth]{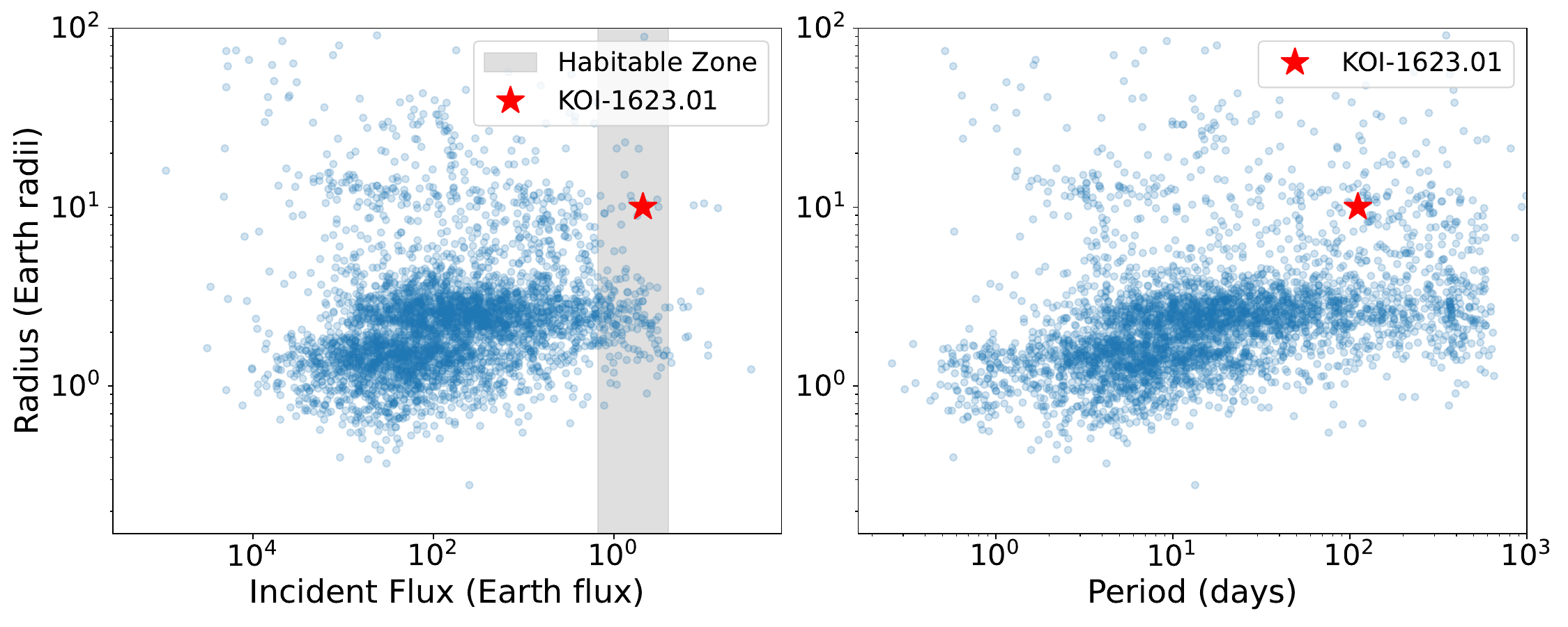}
  \caption{Left: planet incident flux vs. radius for \textit{Kepler} exoplanets. The grey shaded area indicates the approximate ``optimistic'' habitable zone defined by \citet{2016ApJ...830....1K}. Right: planet period vs. radius for \textit{Kepler} exoplanets. The planetary parameters are obtained from \citet{Berger2020AJ....159..280B}. Red star: position of the planet on the diagram as if the companion is the planet host.}
\label{fig:radflux}
\end{figure*}

 Figure~\ref{fig:radflux} shows the planet’s positions on the incident flux–radius and period–radius diagrams along with other \textit{Kepler} systems. If the planet orbits the secondary, a K dwarf, it would be Jupiter-sized, a class of planets that is intrinsically rare, especially around lower-mass stars \citep{Johnson2007ApJ...670..833J,Bonfils2013A&A...549A.109B,Sabotta2021A&A...653A.114S}. \citet{Gaidos2013ApJ...771...18G} report an occurrence rate of $0.7\% \pm 0.5\%$ for giant planets around late K dwarfs in the \textit{Kepler} sample. In this scenario, the planet would reside within the ``optimistic'' habitable zone defined by \citet{2016ApJ...830....1K}, shown as the shaded region in Figure~\ref{fig:radflux}. Previous studies have explored the potential habitability of exomoons orbiting giant planets \citep{Heller2012A&A...545L...8H, Heller2013AsBio..13...18H, Hill2018ApJ...860...67H}. Whether this planet could host such an exomoon requires further investigation.

Two confirmed planets are located in a similar region of the radius–incident flux parameter space. The first is \textit{Kepler}-553c, which was statistically validated by \citet{Morton2016ApJ...822...86M} and follows a moderately eccentric orbit ($e = 0.35$) with an orbital period of 328 days \citep{Dalba2024ApJS..271...16D}.

The second is \textit{Kepler}-1704b, confirmed through radial velocity measurements obtained with the High Resolution Echelle Spectrometer (HIRES) on the Keck I telescope \citep{Dalba2021AJ....162..154D}. It is believed to be a failed hot Jupiter, a giant planet whose periapse is not sufficiently close to its host star to trigger tidal circularization \citep{Dalba2021AJ....162..154D}. \textit{Kepler}-1704b has an extremely eccentric orbit ($e = 0.92$) and a long orbital period of about 1,000 days. By comparison, KOI-1623.01’s shorter orbital period and lower eccentricity make such a scenario unlikely.

\section{Centroid Offset, Not Necessarily False Positives}
\begin{figure*}
  \centering
  \includegraphics[width=0.8\textwidth]{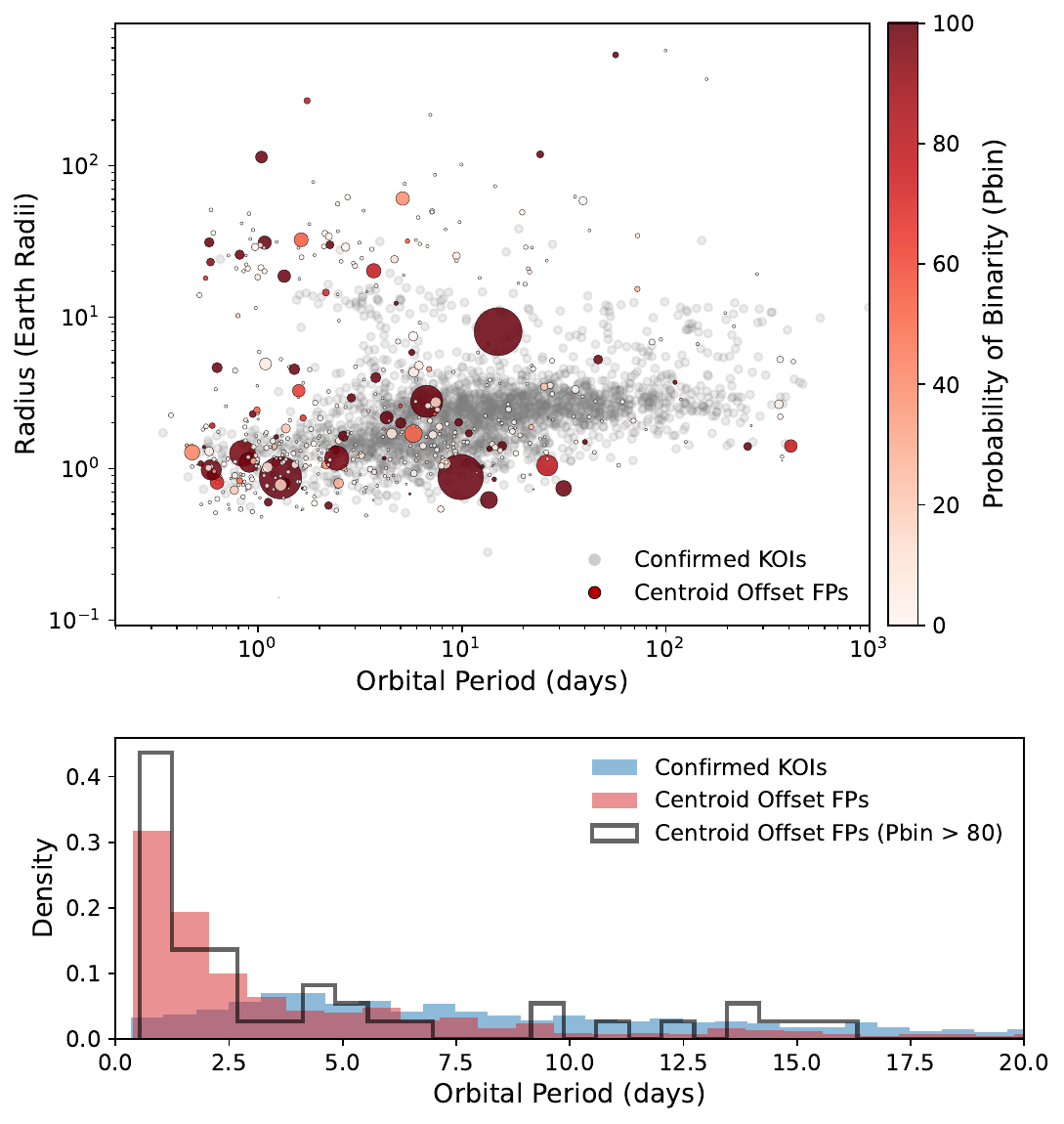}
  \caption{Top panel: Planet radius vs. orbital period for all confirmed KOIs (grey) and centroid-offset false positives (red). Red points are color-coded by the probability that the KOI has a nearby physically bound stellar companion and is therefore more likely to be in a binary system \citep{Pbinpaper}. Point size is inversely proportional to the separation between the binary stars. Bottom panel: Histogram of planet orbital periods for confirmed KOIs (blue), centroid-offset false positives (red), and centroid-offset false positives that are highly likely to be in a binary system ($P_{bin} > 80\%$, black outline).}
\label{fig:discussion}
\end{figure*}

The analysis of KOI-1623.01 implies that planets among the \textit{Kepler} sample may have been misclassified as false positives. Among all KOIs, $\sim$ 500 have been flagged solely as COFPs, with no additional false positive indicators. 

We use binary probabilities ($P_{bin}$) from \citet{Pbinpaper} to perform a quantitative assessment of which COFPs are likely misclassified. The study analyzes nearby stars around all KOIs, estimating the probability that each is a bound binary companion using photometric and astrometric data and distinguishing true binaries from chance alignments. \citet{Pbinpaper} found that among 626 candidate companions to confirmed or candidate hosts with $P_{bin} > 80\%$, only $\sim 13$ are expected to be field interlopers. We therefore adopt $P_{bin} > 80\%$ as the threshold for highly likely bound binaries. Cross-matching this catalog with the list of COFPs, we identify approximately 50 COFPs that have a highly probable binary companion ($P_{bin} > 80\%$). While  high-resolution imaging follow-up has been conducted for many \textit{Kepler} planet candidate host stars \citep[e.g.,][]{Law2014ApJ...791...35L, Baranec2016AJ....152...18B, Ziegler2017AJ....153...66Z}, such observations remain less complete for systems classified as false positives. Many COFPs likely host companions that have yet to be identified, whether bound binaries or chance alignments with field stars. However, it is important to note that some bound binaries may be true astrophysical false positives if the observed transit signal originates from an eclipsing binary on the secondary rather than a planet orbiting the primary. 

Figure~\ref{fig:discussion} presents the radius and orbital period distributions for confirmed KOIs and COFPs. For each COFP, we use Monte Carlo sampling to estimate the number of physically bound systems, finding an average of $\approx$\,70 bound systems over 100 runs. 

The bottom panel of Figure~\ref{fig:discussion} compares orbital periods for confirmed KOIs and COFPs. The confirmed KOIs tend to have longer periods, while COFPs cluster at shorter periods, likely because many COFPs are actual false positives caused by eclipsing binaries with short orbital periods. It is also possible that planets with longer periods and therefore fewer observed transits are less likely to be classified as COFPs, since their centroid offsets are averaged over all quarters and may not reach the significance threshold. We also include the period distribution for COFPs with a high probability of being in binary systems ($P_{bin} > 80\%$). These systems are more prone to misclassification and show peaks at shorter periods ($\sim 1$ day). There are a few possible explanations for the short-period peak. First, because we did not match the binaries’ separations to the centroid offsets, some transits in physically bound binaries may still be false positives caused either by a nearby eclipsing binary or by eclipses occurring within the bound system itself, which often have shorter periods. Second, many of these could be real planets misclassified as false positives, with their period distribution intrinsically different due to the influence of a nearby stellar companion. Several studies have shown that planets in binary systems exhibit a different mass–period distribution compared to those around single stars. In particular, massive short-period planets are found almost exclusively in close binary systems \citep{Zucker2002ApJ...568L.113Z, Eggenberger2004A&A...417..353E, Desidera2007A&A...462..345D, Su2021AJ....162..272S}.

In addition, approximately 30 COFPs have centroid offsets smaller than $1\farcs5$. \textit{Gaia} DR3 completeness declines below $\sim 1\farcs5$ \citep{Gaia2021A&A...649A...1G}, so these close-separation systems are less likely to be resolved by \textit{Gaia} and may not appear in \citet{Pbinpaper}, requiring AO imaging for confirmation. Their small separations suggest they are more likely to be physically bound binaries and host misclassified real planets.

\section{Summary}
In this paper, we analyze KOI-1623.01, previously classified as a centroid offset false positive (COFP), as a planet in an S-type orbit around KOI-1623B. Combining \textit{Kepler} transit photometry and pixel level data, \textit{Gaia} DR3 astrometry, and Keck/NIRC2 imaging, we determine its planetary parameters. Additionally, we assess the potential number of planets that may have been misclassified as COFPs. Our main conclusions are as follows:
\begin{enumerate}
\item{We cross-matched the \textit{Kepler} targets with \textit{Gaia} DR3 and recalculated the centroid offsets using \textit{Gaia} coordinates, KIC coordinates, and the reported KIC centroid offsets. The updated offsets show good agreement with the KIC values, with KIC Offset - \textit{Gaia} Offset having a mean of $-0\farcs0047\pm0\farcs0002$, a standard deviation of $0\farcs084$}.
\item{We performed a pixel-level analysis using \texttt{Transit-APP} by comparing the observed difference image with simulated pixel images under the assumption that the transit originates from either the primary or the secondary star. Model comparison shows that the scenario in which the secondary hosts the transit is strongly preferred.
}
\item{We use \texttt{triceratops} to statistically show that KOI-1623Bb is consistent with a transiting planet. If the planet orbits the secondary K-type star, it would correspond to a giant planet with a radius of $10.020^{+0.437}_{-0.311}\,R_\Earth$, located within the companion star’s habitable zone. However, in the absence of mass constraints, the transit signal on the companion star could also be consistent with a brown dwarf or a low-mass stellar object.}
\item{Among all KOIs, approximately 500 have been flagged solely as COFPs without additional false positive indicators. Using binary probabilities from \citet{Pbinpaper} and Monte Carlo sampling, we estimate $\approx 70$ physically bound systems, which are more likely to host real planets that were misclassified.}
\end{enumerate}
The KOI-1623 system can be followed up with future diffraction-limited spectrographs such as the High-resolution Infrared Spectrograph for Exoplanet Characterization (HISPEC; \citealt{HISPEC2019BAAS...51g.134M}) on the Keck telescopes to obtain precise radial velocity measurements and place constraints on the mass of the transiting companion.

Future studies should further investigate these COFPs. AO imaging of COFP systems would help identify unresolved binaries and quantify the fraction of systems that are currently misclassified. Such efforts will improve our understanding of exoplanet demographics and enhance vetting completeness when calculating planet occurrence rates.

\vspace{0.5cm}
\noindent
The authors wish to recognize and acknowledge the very significant cultural role and reverence that the summit of Maunakea has always had within the Native Hawaiian community. We are most fortunate to have the opportunity to conduct observations from this mountain.

Some of the data presented herein were obtained at Keck Observatory, which is a private 501(c)3 non-profit organization operated as a scientific partnership among the California Institute of Technology, the University of California, and the National Aeronautics and Space Administration. The Observatory was made possible by the generous financial support of the W. M. Keck Foundation. 

R.W. and D.H. acknowledge support from the National Aeronautics and Space Administration (80NSSC22K0781). 

This work made use of the \textit{Gaia}-\textit{Kepler}.fun crossmatch database created by Megan Bedell.

This publication makes use of data products from the Two Micron All Sky Survey, which is a joint project of the University of Massachusetts and the Infrared Processing and Analysis Center/California Institute of Technology, funded by the National Aeronautics and Space Administration and the National Science Foundation.

\textit{Facilities:} NASA Exoplanet Archive, \textit{Kepler}, KeckII, \textit{Gaia}

\textit{Software:} \texttt{exoplanet} \citep{exo2021JOSS....6.3285F}, \texttt{isoclassify} \citep{Huber2017ApJ...844..102H, Berger2023arXiv230111338B, Berger2020AJ....159..280B}, \texttt{ALDERAAN} \citep{umbrella_transit}, \texttt{triceratops} \citep{triceratops, triceratops_code}, \texttt{Lightkurve} \citep{lightkurve}
\clearpage
\appendix
\section{TRICERATOPS Scenario Definitions and Probabilities}

\begin{table*}[htbp]
\centering
\caption{Scenarios tested with \texttt{triceratops} and corresponding mean probabilities, assuming KOI-1623B is the target star.}
\label{tab:triceratops_scenarios}
\begin{tabular}{lll}
\hline
Scenario & Configuration &Mean Probability \\
\hline
TP      & No unresolved companion; transiting planet with $P_{\rm orb}$ around target star & $0.608$ \\
EB      & No unresolved companion; eclipsing binary with $P_{\rm orb}$ around target star & $2.895\times10^{-161}$ \\
EBx2P   & No unresolved companion; eclipsing binary with $2\times P_{\rm orb}$ around target star & $1.333\times10^{-21}$ \\
PTP     & Unresolved bound companion; transiting planet with $P_{\rm orb}$ around target star & $0.140$ \\
PEB     & Unresolved bound companion; eclipsing binary with $P_{\rm orb}$ around target star & $4.855\times10^{-126}$ \\
PEBx2P  & Unresolved bound companion; eclipsing binary with $2\times P_{\rm orb}$ around target star & $2.079\times10^{-22}$ \\
STP     & Unresolved bound companion; transiting planet with $P_{\rm orb}$ around unresolved companion star & $7.485\times10^{-3}$ \\
SEB     & Unresolved bound companion; eclipsing binary with $P_{\rm orb}$ around unresolved companion star & $8.398\times10^{-4}$ \\
SEBx2P  & Unresolved bound companion; eclipsing binary with $2\times P_{\rm orb}$ around unresolved companion star & $1.618\times10^{-23}$ \\
DTP     & Unresolved background star; transiting planet with $P_{\rm orb}$ around target star & $0.244$ \\
DEB     & Unresolved background star; eclipsing binary with $P_{\rm orb}$ around target star & $4.531\times10^{-129}$ \\
DEBx2P  & Unresolved background star; eclipsing binary with $2\times P_{\rm orb}$ around target star & $2.409\times10^{-22}$ \\
BTP     & Unresolved background star; transiting planet with $P_{\rm orb}$ around background star & $1.183\times10^{-5}$ \\
BEB     & Unresolved background star; eclipsing binary with $P_{\rm orb}$ around background star & $3.854\times10^{-11}$ \\
BEBx2P  & Unresolved background star; eclipsing binary with $2\times P_{\rm orb}$ around background star & $7.837\times10^{-11}$ \\
\hline
\end{tabular}
\end{table*}

\clearpage
\bibliography{KOI-1623b}{}
\bibliographystyle{aasjournal}

\end{document}